\documentclass[twocolumn,superscriptaddress,secnumarabic,amssymb,amsmath,nobibnotes,aps,prd,showpacs,nofootinbib]{revtex4}
\usepackage{amsmath}
\usepackage{amsfonts}
\usepackage{amssymb}
\usepackage{graphicx}

\begin{document}
\title{Energy conditions in modified gravity}

\author{Salvatore Capozziello}
\email{capozzie@na.infn.it}
\affiliation{Dipartimento di Fisica, Universit\`{a} di Napoli ``Federico II'', Napoli, Italy
INFN Sez. di Napoli, Compl. Univ. di Monte S. Angelo, Edificio G, Via Cinthia, I-80126, Napoli, Italy.}

\author{Francisco S. N. Lobo}
\email{flobo@cii.fc.ul.pt}\affiliation{Centro de Astronomia
e Astrof\'{\i}sica da Universidade de Lisboa, Campo Grande, Edif\'{i}cio C8
1749-016 Lisboa, Portugal}

\author{Jos\'e P. Mimoso}
\email{jpmimoso@cii.fc.ul.pt}
\affiliation{Centro de Astronomia
e Astrof\'{\i}sica da Universidade de Lisboa, Campo Grande, Edif\'{i}cio C8
1749-016 Lisboa, Portugal}
\affiliation{Departamento de F\'{\i}sica, Faculdade de Ci\^encias da Universidade de Lisboa \\ 
Faculdade de Ci\^encias da Universidade de Lisboa, 
Edif\'{i}cio C8, Campo Grande, P-1749-016 Lisbon, Portugal}

\date{\today}

%\tableofcontents
\begin{abstract}

We consider generalized energy conditions in modified theories of gravity by taking into account the further degrees of freedom related to scalar fields and curvature invariants. The latter are usually recast as generalized {\it geometrical fluids} that have different meanings with respect to the standard matter fluids generally adopted as sources of the field equations. More specifically, in modified gravity the curvature terms are grouped in a tensor $H^{ab}$ and a coupling $g(\Psi^i)$ that can be reorganized  in  effective Einstein field equations, as corrections to the energy-momentum tensor of matter. The formal validity of such inequalities does not assure some basic requirements such as the attractive nature of gravity, so that the energy conditions have to be considered in a wider sense.

\end{abstract}

%\keywords{General Relativity; alternative gravity; cosmology}
\maketitle

%%%%%%%%%%%%%%%%%%%%%%%%%%%%%%%%%%%%%%%%%%%%%%%%%%%%%%%%%%
%%%%%%%%%%%%%%%%%%%%%%%%%%%%%%%%%%%%%%%%%%%%%%%%%%%%%%%%%%
\section{Introduction}
%%%%%%%%%%%%%%%%%%%%%%%%%%%%%%%%%%%%%%%%%%%%%%%%%%%%%%%%%%
%%%%%%%%%%%%%%%%%%%%%%%%%%%%%%%%%%%%%%%%%%%%%%%%%%%%%%%%%%

%{\it Introduction.}
In General Relativity (GR), the Einstein field equations, $G_{ab} = 8\pi G\, T_{ab}$, relate the 
Einstein tensor $G_{ab}\equiv R_{ab}-\frac{1}{2}\,g_{ab}R$,  to the energy-momentum tensor of the matter fields, $T_{ab}$, where $R_{ab}$ is the Ricci tensor, which is defined as the contraction of the Riemann curvature tensor 
${R^c}{}_{acb}=R_{ab}$, and $R={R^a}_a$ is the curvature scalar. The Einstein equations govern the interplay between the geometry of the spacetime and the matter content. There is a clear separation between the left-hand side  that corresponds to the geometry, and the right-hand side where one finds the energy-matter distribution. The underlying idea is that the matter-energy distribution tell us how the spacetime is curved and, hence, how gravity acts. Therefore, it follows from the equations, that any conditions that we impose on $T_{ab}$ immediately translate into corresponding conditions on the Einstein tensor $G_{ab}$ \cite{Hawking:1973uf}. In this sense, the causal and geodesic  structures of space-time are determined by the matter-energy distribution. In this context, the energy conditions guarantee that the causality principle is respected and suitable physical {\it sources} have to be considered \cite{Hawking:1973uf,Wald}.

The definition of the energy conditions entails an arbitrary flow which represents a generic observer or a reference frame. In general, we consider a congruence of timelike curves whose tangent 4-vector $W^a$ represents the velocity vector of a family of observers. Alternatively, we may consider a field of null vectors, $k^a$, which has the advantage in simplifying $G_{ab}k^a k^b =R_ {ab}k^ak^b$, since $g_{ab}\, k^ak^b=0$ by assumption. Thus, the energy conditions emerge directly from the geodesic structure of the space-time. More specifically, consider the {\it Raychaudhuri equation}, given by \cite{falco}
\begin{equation}
\dot \theta + \frac{\theta^2}{3}+2\,(\sigma^2-\omega^2) - \dot{W}^a{}_{;a}=- R_ {ab} \, W^a W^b  \;. \label{Ray_1}
%\dot \theta + \frac{\theta^2}{3}+2\,(\sigma^2-\omega^2) - \dot{u}^a_{;a}= -4\pi G\, (\rho+3p) \; ,
\end{equation}
where $\sigma_{ab}$ is the {\it shear tensor}, $\theta$ is the {\it expansion scalar} and $\omega_{ab}$ is the vorticity tensor.
It is important to emphasize that Eq. (\ref{Ray_1})  carries only  a geometrical meaning, as the quantities in it are  directly derived from the Ricci identities. It is only when we choose a particular theory that we establish a relation between $R_ {ab} \, W^a W^b $ in Eq. (\ref{Ray_1}), and the energy-momentum tensor describing matter fields \cite{Hawking:1973uf,Wald}. One may also consider a null congruence $k^a$ and a vanishing  vorticity $\omega_{ab}=0$, which means that, in GR, it is possible to associate  the null energy condition with the focusing (attracting) characteristic of the spacetime geometry.

In this work, we tackle the problem of the energy conditions in modified gravity. This issue is extremely delicate since a standard approach is to consider the gravitational field equations as  effective Einstein equations. More specifically, the further degrees of freedom carried by these theories \cite{Hehl:1976,Hehl:1995,Trautman:2006} can be recast as generalized {\it geometrical fluids} that have different meanings with respect to the standard matter fluids generally adopted as sources of the field equations \cite{report}. While standard fluids generally obey standard equations of states, these ``fictitious'' fluids can be related to scalar fields or further gravitational degrees of freedom, as in $f(R)$ gravity. In these cases, the physical properties may be ill-defined and the consequences can be dramatic, since the causal and geodesic structures of the theory could present  serious shortcomings as well as the energy-momentum tensor could not be consistent with the Bianchi identities and the conservation laws.
Thus, we add a cautionary note of the results obtained in the literature \cite{ETG_ec}.
Finally, we have to mention other  important resuts for energy conditions in alternative gravity. For example, in  \cite{atazadeh},  energy conditions in $f(R)$ gravity and Brans-Dicke theories are discussed. In \cite{albareti},  the non-attractive character of gravity in $f(R)$ theories is considered while energy conditions in the Jordan frame are taken into account in \cite{saugata}. 
In this work, we adopt the $(-+++)$ signature and $c=1$.

\section{Energy conditions in modified theories of gravity}

%{\it Energy conditions in modified theories of gravity.}
In the context of modified theories of gravity, at least for a large class of interesting cases, the generalized field equations can be cast in the following form
\begin{equation}
g(\Psi^i)\, \left( G_{ab} +H_ {ab}\right)= 8\pi G\, T_{ab} \; , \label{ETG_EFE1}
\end{equation}
where $H_{ab}$ is an additional geometrical term with regard to GR that encapsulates the geometrical modifications introduced by the modified theory, and $g(\Psi^i)$ is a factor that modifies the coupling with the matter fields in $T^{ab}$, where $\Psi^i$ generically represents either curvature invariants or other gravitational fields contributing to the dynamics. GR is recovered for $H_{ab}=0$ and $g(\Psi^i)=1$.

Taking into account the diffeomorphism invariance of the matter action,  the covariant conservation of the 
energy-momentum tensor, $\nabla_a T^{ab}=0$, is obtained. Thus, from the contracted Bianchi identities, we derive the following conservation law
\begin{equation}
\nabla_b H^{ab} = - \frac{8\pi G}{g^2} T^{ab}\, \nabla_b g\; .
\end{equation}
The fact that $H^{ab}$ is a geometrical quantity, in the sense that it can be given by geometrical invariants or scalar fields different from ordinary matter fields, implies that the imposition of a specific energy condition on $T^{ab}$ carries an implication for the  combination of $G_ {ab}$ with $H_ {ab}$ and not just for the Einstein tensor.  So we cannot obtain a simple geometrical implication, as in GR, from it any more. For instance, if we assume that the strong energy condition, $T_{ab}\, W^a W^b  \ge \frac{1}{2}T\,W^aW_a$ holds, it would mean, on the one hand, in GR that $R_ {ab} \, W^a W^b  \ge 0$ and, on the other hand, given Eq. (\ref{Ray_1}), that the geodesics are focusing, and hence that gravity possesses an attractive character. This is one of the assumptions of the singularity theorems of Hawking and Penrose \cite{Hawking:1973uf}. However in the modified gravity context under consideration,  this condition just states that 
\begin{equation}
g(\Psi^i)\,  \left(R_{ab}+H_ {ab}-\frac{1}{2}g_{ab} H \right)\,W^aW^b \ge 0\; , \label{en_cond_strong_2}
\end{equation}
which does not necessarily imply $R_ {ab} \, W^a W^b  \ge 0$ and hence we cannot straightforwardly conclude that the satisfaction of the strong energy condition (SEC) is synonymous of the attractive nature of gravity in the particular modified theory of gravity under consideration.

The term $H^{ab}$ is usually treated, in the literature, as a correction to the energy-momentum tensor, so that the meaning which  is attributed to the energy conditions is the satisfaction of a specific inequality using the combined quantity $T^{ab}_{\rm eff}=T^{ab}/g-H^{ab}$.  It is thus misleading to associate this effective 
energy-momentum tensor to the energy conditions, since they do not emerge only from $T^{ab}$ but from the geometrical quantity $H^{ab}$, which is considered as an  additional energy-momentum tensor. 

However, if the modified theory of gravity under consideration allows an equivalent description  upon an appropriate conformal transformation, it then becomes justified to associate the transformed $H^{ab}$ to the redefined $T^{ab}$ in the conformally transformed  Einstein frame. In fact, conformal transformations play an extremely relevant role in the discussion of the energy conditions. In particular, they allow to emphasize the further degrees of freedom coming from modified gravities under the form of curvature invariants and scalar fields. Specifically, several   generalized  theories of gravity  can be redefined as GR plus a number of appropriate fields coupled to matter by means of a conformal transformation in the so-called Einstein frame. This is, for instance, the case for scalar-tensor gravity theories, for $f(R)$ gravity, etc \cite{report}. 

Indeed, in the scalar-tensor case, although in the Jordan frame one has a separation between  geometrical terms and  standard matter terms that can be cast as in (\ref{ETG_EFE1}),  where $H_{ab}$  involves a mixture of both the scalar and tensor gravitational fields, i.e., of $\varphi$ and $R^{ab}, R$, it happens that upon a suitable conformal transformation we are able to cast the field equations as
%\begin{equation}
$\tilde{G}_ {ab}= 8\pi G\, \tilde{T}_{ab}^{\rm eff}$, 
%\label{STT_EF}
%\end{equation}
where $\tilde{T}_{ab}^{\rm eff}=\tilde{T}^M_{ab}+\tilde{T}^\varphi_{ab}$. It thus makes sense to consider $\tilde{T}_{ab}^{\rm eff}$ as an effective energy-momentum tensor, where $\tilde{T}^M_{ab}$ is the transformed energy-momentum of matter, and $\tilde{T}^\varphi_{ab}$ is an energy-momentum tensor for the redefined scalar field $\varphi$ which is coupled to matter. 
Then one finds results where one draws conclusions about the properties of $\tilde{G}_ {ab}$ such whether it focuses geodesics directly from those conditions holding on $\tilde{T}_{ab}^{\rm eff}$. This ignores the fact that $H_{ab}$ originally possesses a geometrical character, and thus the conclusions may be too hasty if not supported by the physical analysis of the sources. 

If we assume that in this frame the effective energy-momentum tensor $ \tilde{T}^{\rm eff}_{ab}$ satisfies some energy condition, for instance, the null energy condition (NEC), this implies that ${\tilde G}_{ab}$ has to satisfy such a condition. Thus, it is possible to write the Raychaudhuri equation as
\begin{equation}
\frac{{\rm d} \tilde\theta}{{\rm d}v} = - \left[\frac{\tilde\theta^2}{3}+2\tilde\sigma^2 +\tilde R_{ab}\tilde k^a\tilde k^b\right]\; , \label{eq_null}
\end{equation}
which enables us to conclude on the attractive/repulsive character of the given theory of gravity in the Einstein frame. Reversing the conformal transformation,  we can assess, in principle,  what happens in the original frame, namely, the Jordan frame. This  operation  requires to  know   how the kinematical quantities, present in 
Eq. (\ref{eq_null}), transform under a conformal transformation.
This means that if $g_{ab}\to \tilde{g}_{ab}=\Omega^2\,g_{ab}$ and $W^a \to \tilde W^a =\Omega^{-1} W^a$, we have
%\begin{equation}
$\tilde \nabla_a \tilde W_b = \Omega \, \nabla_a W_b +\Omega \, {\gamma^c}_{ab} W_c + W_b\,\nabla_a\Omega $, 
%\end{equation}
where 
%\begin{equation}
${\gamma^c}_{ab} = \delta^c_a \partial_b\Omega/\Omega+\delta^c_b \partial_a\Omega/\Omega-g_{ab} \partial^c\Omega/\Omega $.
%\; .
%\end{equation}

From this result, it follows that we can pass from the Einstein to the Jordan frame by the following transformations
$\tilde{\theta}_{ab} = \Omega\, ( {\theta}_{ab} - \dot\Omega\, h_{ab} )$,
$\tilde{\sigma}_{ab} =  \Omega\,\sigma_{ab} $,
$\tilde\omega_{ab} = \Omega\,\omega_{ab} $,
$\tilde{\theta} = \Omega^{-1}\,( \theta-3\dot\Omega )$,
respectively.
Thus, Eq. (\ref{eq_null}) can finally be written as
%\begin{equation}
$\frac{{\rm d} \tilde\theta}{{\rm d}v}= \frac{\dot\theta}{\Omega^2}-\frac{\theta}{\Omega^2}\frac{\dot\Omega}{\Omega} -\frac{3}{\Omega}\left(\ln \Omega\right)^{..}$.
%\end{equation}
%
The latter result shows that whereas, in the Einstein frame,  the NEC implies the attractive nature of gravity, a similar implication does not necessarily follow in  the Jordan frame. In fact, $d\tilde\theta/dv\le 0$ only implies that
%\begin{equation}
$\dot\theta \le \frac{\dot\Omega}{\Omega} \theta+{3\Omega}\left(\ln \Omega\right)^{..} $,
%\end{equation}
and thus it depends on the sign of the term on the right-hand side of the inequality. On the other hand, we see that $\tilde R_{ab}\tilde k^a\tilde k^b\ge 0$ does not necessarily entail
$ R_{ab}k^ak^b \ge 0$. What we do indeed obtain is
\begin{equation}
\left( \Omega^{-2} R_{ab}+   2\nabla_a\nabla_b\ln \Omega +2\nabla_a\ln\Omega\,\nabla_b\ln\Omega \right)\,k^a k^b  \ge 0 \,.
\end{equation}
%if $\tilde R_{ab}\tilde k^a\tilde k^b\ge 0$.

This discussion emphasizes that if, for example, in one of the conformally related frames, we have  attractive gravity (due to the NEC), in the other frame neither the NEC is simultaneously satisfied, nor, in case it is, this  means that gravity will be straightforwardly attractive. This fact could be extremely relevant in view of identifying a physical meaning of conformal transformations.  The debate could be fixed as soon as a set of conformally invariant physical quantities is identified. However, some physical quantities, like mass, are not conformally invariant so some authors claim that such  transformations are just a mathematical tool to change  frames, while others argue that conformal transformations have a true physical meaning \cite{report,basilakos}. As discussed in \cite{magnano} for $f(R)$ gravity, the energy conditions could greatly aid in this debate.  
We refer the reader to \cite{Catena:2006bd} for a discussion on the formulation of scalar-tensor theories of gravity in the Einstein and the Jordan frames in a cosmological context. 

%%%%%%%%%%%%%%%%%%%%%%%%%%%%%%%%%%%%%%%%%%%%%%%%%%%%%%%%%%
%%%%%%%%%%%%%%%%%%%%%%%%%%%%%%%%%%%%%%%%%%%%%%%%%%%%%%%%%%
\section{Example of a modified theory of gravity: Scalar-tensor gravity}\label{secIV}
%%%%%%%%%%%%%%%%%%%%%%%%%%%%%%%%%%%%%%%%%%%%%%%%%%%%%%%%%%
%%%%%%%%%%%%%%%%%%%%%%%%%%%%%%%%%%%%%%%%%%%%%%%%%%%%%%%%%%

%{\it Example of a modified theory of gravity: Scalar-tensor gravity.}
According to the above discussion, the possibility to formulate the energy conditions for any modified gravity strictly depend on the correct identification of the function $g(\Psi^i)$, related to the gravitational coupling, and the tensor $H_{ab}$, which contains the further degrees of freedom of the theory with respect to GR. 
%The former is related to the gravitational coupling that can be non-minimal, the latter is the contribution to the %effective energy-momentum tensor containing the further degrees of freedom of the theory with respect to GR.

Consider scalar-tensor gravity~\cite{ST} given by the action
\begin{equation}
S=\frac{1}{16\pi} \int \sqrt{-g} d^4x\, \left[\phi R - \frac{\omega(\phi)}{\phi}
\phi_{,a} \phi^{,a} + 2 \phi \lambda(\phi)\right] + S_M \,,
\end{equation}
where $S_M$ is the standard matter part, the gravitational coupling is assumed variable and a self-interaction potential is present. 
Varying this action with respect to the metric $g_{ab}$  and the scalar field $\phi$ yields the field equations (\ref{ETG_EFE1}), with $H_{ab}$ given by
\begin{eqnarray}
H_{ab}&=&-\frac{\omega(\phi)}{\phi^2}\;
\left[\phi_{;a}\phi_{;b} - \frac{1}{2} \, g_{ab}\, 
\phi_{;c}\phi^{;c}\right]
   \nonumber \\
&& - \frac{1}{\phi} \left[\phi_{;ab}-g_{ab}
{\phi_{;c}}^{;c}\right]-\lambda(\phi)g_{ab} \,, \label{H-ab_ST}
\end{eqnarray}
and $g(\Psi^i)=\phi$, which we shall assume positive, and
\begin{eqnarray}
&&\Box{\phi}+\frac{{2\phi^2\lambda'(\phi)-2\phi\lambda(\phi)}}
{{2\omega(\phi)+3}}
   \nonumber \\
&&= \frac{1}{2\omega(\phi)+3}\;\left[ 8\pi G\, T-\omega'(\phi) 
\phi_{;c}\phi^{;c}
\right] \,,
\end{eqnarray}
where $T\equiv T^{c}{}_{c}$ is the trace of the matter energy-momentum tensor and $G \equiv\frac{2\omega+4}{2\omega+3}\,$ is the gravitational constant normalized to the Newton value. One also requires  the conservation of the matter content $\nabla^a T_{ab}=0$, to preserve  the  equivalence principle. The archetype Brans-Dicke theory is characterized by the restriction of $\omega(\phi)$ being a constant, and of $\lambda=\lambda'=0$. 

The above considerations on the energy conditions apply straightforwardly. In particular, 
Eq. (\ref{en_cond_strong_2}) is easily recovered  like the other energy conditions. 
Since we assume $\phi>0$, we see that the condition $R_{ab}\,W^a\,W^b\ge 0 $ yielding the focusing of the time-like congruence, and hence attractive gravity, becomes 
\begin{eqnarray}
( T_{ab}  - \frac{1}{2}\,g_{ab}\,T)\,W^aW^b \ge \phi\,( H_{ab}  - \frac{1}{2}\,g_{ab}\,H)\,W^aW^b . \label{en_cond_strong_ST1}
\end{eqnarray}
%We thus realize that the usual SEC associated with attractive gravitation requires that
%$ H_{ab}  - \frac{1}{2}\,g_{ab}\,H \le 0 $.
We further notice that the satisfaction of the latter condition  allows for the focusing of the time-like paths even when a mild  violation of the energy condition occurs. This is an interesting result since matter may exhibit unusual thermodynamical features, e.g. including negative pressures, and yet gravity remains attractive. Alternatively, we see that repulsive gravity may occur for common matter, i.e., for matter that satisfies all the energy conditions (see \cite{bamba1}). 
%This happens when  $H_{ab}$ has the reverse sign in (\ref{en_cond_strong_ST1_b}).

Indeed, the inequality (\ref{en_cond_strong_2}) may be expressed as 
\begin{eqnarray}
&W^aW^b  \Big[ \frac{8\pi}{\phi}\,\left(T_{ab}-\frac{\omega+1}{2\omega+3}\,g_{ab}\,T\right) 
+\frac{\omega}{\phi^2}\nabla_a\phi\nabla_b \phi+\frac{\nabla_a\nabla_b \phi}{\phi}  
     \nonumber \\   
&-\frac{1}{2\phi}\frac{\omega'}{2\omega+3}\,g_{ab}\nabla_c\nabla^c \phi - g_{ab}\,\frac{\phi\lambda'+(\omega+1)\lambda}{2\omega+3}\Big] \ge 0  \; . \label{en_cond_strong_ST4}
\end{eqnarray}
If we consider a Friedmann-Lema\^{\i}tre-Robertson-Walker universe (FLRW) we derive
\begin{equation}
\frac{8\pi G}{\phi}\,\frac{(\omega+3)\rho+3\omega p}{2\omega+3} +\frac{\lambda}{3} + \frac{\omega}{3}\,\frac{\dot\phi^2}{\phi^2}+ \frac{\dot\omega}{2(2\omega+3)}\,\frac{\dot\phi}{\phi} +H\frac{\dot\phi}{\phi}\ge 0 \; ,
\end{equation} 
where the functions $\omega(\phi)$ and $\lambda(\phi)$ clearly define whether gravity is attractive or repulsive.

However, it is interesting to note that, in close analogy with the decomposition of the energy-momentum tensor with respect to the vector field $W^a$ \cite{Wald,Hawking:1973uf}, one may consider the following useful geometrical quantities
\begin{eqnarray}
\tilde\rho =g H_{||}= (gH_{ab})\,W^aW^b, \label{redef_tilderho}  
\;\;\,
3\tilde p=3gH_{\bot} =(gH_{ab})\,h^{ab}, \;
%\label{redef_tilderho-b} 
  \\
\tilde\Pi^{ab} = gH_{\bot}^{<ab>}= \left(h^{ac}h^{bd}-\frac{1}{3}h^{ab}h^{cd}\right)(g\,H_{cd}), \; \label{redef_tilderho-c} 
   \\
\tilde q^a = gH_{\bot}^{a}= W^c \,(g\,H_{cd})\,h^{ad} , \;\, \label{redef_tilderho-d}
\end{eqnarray} 
where $H_{||}$ and $H_{\bot}$ are scalars, $H_{\bot}^{a} $ is a vector and $H_{\bot}^{<ab>}$ is a projected trace-free symmetric tensor.

The decomposition (\ref{redef_tilderho})--(\ref{redef_tilderho-d}) of the tensor $H_{ab}$ into the parallel and orthogonal components to the time-like vector flow $W^a$ is given by 
\begin{eqnarray}
H^{ab} &=& H_{||} W^aW^b + H_{\bot} h^{ab} + 2\,H_{\bot}^{(a}\, W^{b)} + H_{\bot}^{<ab>}
  \nonumber   \\
&=& \frac{1}{\phi}\,\left[  \tilde\rho W^aW^b + \tilde p h^{ab} + 2\,\tilde q^{(a}\, W^{b)} + \tilde\pi^{ab}\right] \,.
\end{eqnarray} 
Thus, the inequality (\ref{en_cond_strong_ST1}) may be written as
%\begin{equation}
$(\rho+3p)/\phi-(H_{||}+3H_{\bot}) \ge 0$,
%\end{equation}
where we have  used the definitions
\begin{eqnarray}
H_{||} &=&  - \frac{\omega(\phi)}{2\phi^2}\,\left(3\dot\phi^2-h^{cd}\,\nabla_c\phi\, \nabla_c\phi\right) \nonumber \\ 
&& -\frac{1}{\phi}\,h^{cd}\nabla_c\nabla_d\phi +\lambda(\phi) \,,
\end{eqnarray}
\begin{eqnarray}
H_{\bot} &=& - \frac{\omega(\phi)}{3\phi^2}\,\left( \frac{\dot\phi^2}{2}-\frac{1}{2}h^{cd}\,\nabla_c\phi\, \nabla_c\phi\right)  \nonumber \\ 
&& \hspace{-1cm} -  \frac{1}{2\phi}\,\left(W^aW^b \, \nabla_c\nabla_d\phi -\frac{1}{3}\,h^{cd}\nabla_c\nabla_d\phi\right) - \lambda(\phi). 
\end{eqnarray}
Thus, $\omega(\phi)$ and $\lambda(\phi)$ define whether gravity is attractive or repulsive in the scalar-tensor cosmological models.
On the other hand, upon conformally transforming the theory into the Einstein frame by $g_{ab}\to \bar g_{ab} = (\phi/\phi_\ast)\, g_{ab}$, the  condition for gravity to be attractive  with the redefined Ricci tensor becomes
\begin{equation}
\tilde R_{ab} \tilde{W}^a \tilde{W}^b= \frac{4\pi}{\phi_\ast} \, (\bar\rho+3\bar p) + \frac{8\pi}{\phi_\ast}\, \left[\dot{\varphi}^2-\tilde{V}(\varphi)\right] \ge 0 \,,
\end{equation}
where $\varphi= \int \sqrt{(2\omega+3)/2}\; {\rm d}\ln\phi$ is the redefined scalar field,  $V(\varphi)= \lambda(\phi(\varphi))/\phi(\varphi)$ is the rescaled potential, and $\bar\rho = \rho/\phi^2$, $\bar{p}=p/\phi^2$. So, although the latter condition  adopts the familiar form found in GR models endowed with a combination of matter and a scalar field, the role of the  functions $\omega(\phi)$ and $\lambda(\phi)$ underlies the result because the definitions of $\varphi$ and $V(\varphi)$ depend on them. In addition, in the Einstein frame, the matter and the scalar field are interacting with each other as revealed by the scalar field equation
\begin{equation}
\ddot\varphi+\bar{\theta}\dot\varphi= -\frac{\partial V(\varphi)}{\partial \varphi}- \frac{\partial \bar\rho(\varphi,\bar{a})}{\partial \varphi}\; .
\end{equation}
Thus, the dependence of the self-interacting potential $V(\varphi)$, and the coupling $\partial_\varphi \bar{\rho}\propto \alpha(\varphi) a^{-3\gamma}$ is important, where $\alpha=(\sqrt{2\omega+3})^{-1}$, when considering a  perfect fluid with $\bar p=(\gamma-1)\bar\rho$. In a cosmological setting, the interplay of the intervening components such that those which violate the SEC dominate imply that  gravity  exhibits a transition from being attractive into becoming repulsive. This feature is relevant in view of dark energy.
%
%The typical case is provided when $V(\varphi)$ has a non-vanishing minimum\cite{ETG_2_GR}.

%%%%%%%%%%%%%%%%%%%%%%%%%%%%%%%%%%%%%%%%%%%%%%%%%%%%%%%%%%
%%%%%%%%%%%%%%%%%%%%%%%%%%%%%%%%%%%%%%%%%%%%%%%%%%%%%%%%%%
\section{Discussion and conclusions}\label{sec:concl}
%%%%%%%%%%%%%%%%%%%%%%%%%%%%%%%%%%%%%%%%%%%%%%%%%%%%%%%%%%
%%%%%%%%%%%%%%%%%%%%%%%%%%%%%%%%%%%%%%%%%%%%%%%%%%%%%%%%%%

%{\it Discussion and conclusions.}
In this paper, we have discussed the formulation and the meaning of the energy conditions in the context of modified theories of gravity. The procedure consists  in  disentangling the further degrees of freedom that emerges with respect to GR and in grouping them as an effective energy-momentum tensor of the form $T^{ab}/g-H^{ab}$ where $g(\Psi^i)$ is the effective coupling and $H^{ab}$ the contribution due to scalar fields and/or curvature invariants of the given modified theory of gravity. 
Formally, the weak, null, dominant and strong energy conditions can be rewritten as in GR. Despite of this analogy, their meaning can be totally  different with respect to GR since the causal structure, geodesic structure and gravitational interaction may be altered.
  
A main role in this analysis is played by recasting the theory, by conformal transformations,  in the Einstein frame where matter and geometrical quantities can be formally dealt exactly such as in GR.  However, the energy conditions can assume a completely different meaning going back to the Jordan frame and then they could play a crucial role in identifying the physical frame as firstly pointed out in \cite{magnano}. 
On the other hand, geometrical implications change in the two frames since optical scalars like $\sigma$, $\theta$ and $\omega$ can give rise to the convergence or divergence of geodesics. This means that the physical meaning of a given extended theory strictly depends on the energy conditions and initial conditions (in relation to the choice of the source \cite{vignolo}). From an observational point of view, this fact could constitute a formidable tool to test the dark components since deviations from standard GR could be put in evidence.

%%%%%%%%%%%%%%%%%%%%%%%%%%%%%%%%%%%%%%%%%%%%%%%%%%%%%%%%%%
\section*{Acknowledgements}
%%%%%%%%%%%%%%%%%%%%%%%%%%%%%%%%%%%%%%%%%%%%%%%%%%%%%%%%%%

%{\it Acknowledgements.}
SC acknowledges the INFN ({\it iniziative specifiche} TEONGRAV and QGSKY). FSNL is supported by a Funda\c{c}\~{a}o para a Ci\^{e}ncia e Tecnologia Investigador FCT Research contract, with reference IF/00859/2012, funded by FCT/MCTES (Portugal). FSNL and JPM acknowledge financial support of the Funda\c{c}\~{a}o para a Ci\^{e}ncia e Tecnologia through the grants CERN/FP/123615/2011 and CERN/FP/123618/2011.

%\newpage
%%%%%%%%%%%%%%%%%%%%%%%%%%%%%%%%%%%%%%%%%%%%%%%%%%%%%%%%%%

\end{document}